# *AdaComm*: Tracing Channel Dynamics for Reliable Cross-Technology Communication

Weiguo Wang[1], Xiaolong Zheng[2], Yuan He[1], Xiuzhen Guo[1]
[1]School of Software and BNRist, Tsinghua University
[2]School of Computer Science, Beijing University of Posts and Telecommunications
wwg18@mails.tsinghua.edu.cn, zhengxiaolong@bupt.edu.cn,
he@greenorbs.com, guoxz16@mails.tsinghua.edu.cn

*Abstract*—Cross-Technology Communication (CTC) is an emerging technology to support direct communication between wireless devices that follow different standards. In spite of the many different proposals from the community to enable CTC, the performance aspect of CTC is an equally important problem but has seldom been studied before. We find this problem is extremely challenging, due to the following reasons: on one hand, a link for CTC is essentially different from a conventional wireless link. The conventional link indicators like RSSI (received signal strength indicator) and SNR (signal to noise ratio) cannot be used to directly characterize a CTC link. On the other hand, the indirect indicators like PER (packet error rate), which is adopted by many existing CTC proposals, cannot capture the short-term link behavior. As a result, the existing CTC proposals fail to keep reliable performance under dynamic channel conditions. In order to address the above challenge, we in this paper propose AdaComm, a generic framework to achieve self-adaptive CTC in dynamic channels. Instead of reactively adjusting the CTC sender, AdaComm adopts online learning mechanism to adaptively adjust the decoding model at the CTC receiver. The self-adaptive decoding model automatically learns the effective features directly from the raw received signals that are embedded with the current channel state. With the lossless channel information, AdaComm further adopts the fine tuning and full training modes to cope with the continuous and abrupt channel dynamics. We implement AdaComm and integrate it with two existing CTC approaches that respectively employ CSI (channel state information) and RSSI as the information carrier. The evaluation results demonstrate that AdaComm can significantly reduce the SER (symbol error rate) by 72.9% and 49.2%, respectively, compared with the existing approaches.

## I. Introduction

The ever-developing Internet of Things (IoT) brings the widespread deployments as well as the rich diversity of wireless technologies [1]–[3]. To directly interconnect the heterogeneous devices that follow different wireless technologies, Cross-Technology Communication (CTC) is proposed to enable the direct communication between incompatible devices without extra hardware.

Despite the tremendous advances, existing CTC approaches usually focus on enabling the communication between incompatible technologies. How to maintain the reliable performance in the intrinsically dynamic channels has not received enough attention. To convey data, existing CTC techniques usually explore the mutually accessible information carrier such as the energy and timing of packet transmissions [4]–[7], the state variations of overlapped channels [8], [9], and the originally incompatible but similar signals [10]–[12]. Since most CTC leverages the signal patterns rather than the underlying raw signals to convey data, the CTC links significantly differ from the links in traditional wireless communication. Traditional link quality indicators such as RSSI or CSI used by WiFi to learn the channel state cannot reflect the quality of a CTC link. So far there is not a link quality indicator to accurately describe the CTC's channel state.

Without an accurate CTC link indicator to learn the current channel state, to cope with channel dynamics is a fundamental but challenging task for CTC. Most of the existing CTC approaches adopt indirect indicators, e.g. Packet Error Rate (PER), to detect the changes of channel state. When there is a significant variation in the PER, the CTC approaches may reactively control the sender's encoding behavior, so as to enhance the features of encoded signals received by the receiver. For example, WiZig [6] extends the symbol window length and enlarges the differences of adjacent encoded amplitudes to enhance the features of CTC symbols. FreeBee [4] increases the number of beacon repetitions per symbol in the noisy channel, thus improving the highest fold sum. ZigFi [8] controls the transmission power to maintain Signal-to-Interference-plus-Noise-Ratio (SINR) perceived by the receiver so that the CSI of ZigFi symbols still satisfy the decoding model.

The reactive adjustments of CTC against channel dynamics, unfortunately, suffer performance degradation and even failure in practice. First, the indirect indicators of channel state only reflect the long-term average channel quality but are insensitive to the short-time channel dynamics. Therefore, adjusting CTC according to those indicators can only achieve sub-optimal performance. Second, existing decoding methods usually adopt the threshold or machine learning model as the pattern recognition methods. But features such as the RSSI threshold and CSI variation predefined by the feature-based decoding model cannot accurately describe the channel dynamics. For example, the PHY information such as CSI can be affected by both the intentional CTC transmissions and the channel-related factors like multipath. The predetermined statistical features are not necessarily effective to cover all possible cases, due to the uncertainty of channel dynamics.

In order to address the above problems, in this paper we propose AdaComm, a general and lightweight online adaptive

CTC framework that automatically adjusts the decoding model to maintain reliable communication performance in dynamic channel conditions. Instead of enhancing the features of signals from the sender, we propose a self-adapting decoding model at the receiver side, which traces the channel state to improve the decoding reliability. We directly use raw received data as input and avoid the information loss caused by manual feature extraction. Our model automatically extracts the effective features to distinguish between the intended impact of CTC modulation and the channel dynamics. We also design an online learning mechanism that leverages the correctly decoded CTC data to update the decoding model without extra cost of data collection, which is called fine tuning. With fine tuning, AdaComm is able to cope with continuous changes of the channel state. To deal with model failures caused by abrupt channel changes, AdaComm integrates a full training mode that retrains the decoding model with the newly collected training sequences. To reduce the cost of data collection for full training, we devise a data augmentation method to obtain sufficient training data with only limited size of the training sequence. The main contributions of this work are summarized as follows.

- We propose AdaComm, a general online learning CTC framework to maintain reliable performance in dynamic channels. AdaComm integrates a lightweight decoding model that takes the information of channel state into consideration and automatically extracts decoding features.
- We design fine tuning and full training modes to cope with continuous and abrupt channel changes. In fine tuning, we use the correctly decoded CTC data to update the decoding model. In full training, we propose a data augmentation method to reduce the cost of collecting online training data.
- We implement AdaComm on both CSI-based and RSSI-based CTC and evaluate its performance in various environments. The experiment results show that AdaComm can reduce the SER by 72.9% and 49.2% for CSI-based and RSSI-based CTC, respectively.

The rest of this paper is organized as follows. We present the related work in Section II. In Section III, we investigate the performance of existing CTC in a dynamic environment and analyze the causes of performance degradation. Section IV presents the design of AdaComm. We evaluate the performance of AdaComm in Section V and conclude our work in Section VI.

## II. RELATED WORK

Cross-Technology Communication (CTC) has been developing rapidly and applied for channel coordination among heterogenous technologies [13], [14]. The common idea of CTC is building the mutually accessible information carrier with existing hardware to convey data. One of the most common information carriers is the energy of packet transmissions. ESense [5] modulates symbols by packet lengths and accomplishes CTC from WiFi to ZigBee. HoWiEs [15] improves Esense by using combinations of WiFi packets.

| | | Validation Dataset | | | | | | | | |
|---|---|---|---|---|---|---|---|---|---|---|
| Time | 12:30(0) | 12:45(1) | 13:00(2) | 13:15(3) | 13:30(4) | 13:45(5) | 14:00(6) | 14:15(7) | 14:30(8) | 14:45(9) |
| 12:30(0) | 0.913 | 0.611 | 0.544 | 0.755 | 0.429 | 0.587 | 0.543 | 0.507 | 0.551 | 0.474 |
| 12:45(1) | 0.833 | 0.956 | 0.826 | 0.911 | 0.929 | 0.809 | 0.670 | 0.503 | 0.524 | 0.564 |
| 13:00(2) | 0.854 | 0.942 | 0.930 | 0.770 | 0.790 | 0.901 | 0.937 | 0.720 | 0.604 | 0.568 |
| 13:15(3) | 0.673 | 0.404 | 0.888 | 0.921 | 0.936 | 0.901 | 0.589 | 0.562 | 0.660 | 0.719 |
| 13:30(4) | 0.753 | 0.611 | 0.946 | 0.923 | 0.960 | 0.717 | 0.584 | 0.541 | 0.668 | 0.761 |
| 13:45(5) | 0.611 | 0.620 | 0.515 | 0.895 | 0.895 | 0.914 | 0.918 | 0.795 | 0.741 | 0.683 |
| 14:00(6) | 0.630 | 0.607 | 0.548 | 0.869 | 0.607 | 0.710 | 0.956 | 0.904 | 0.825 | 0.793 |
| 14:15(7) | 0.580 | 0.480 | 0.485 | 0.728 | 0.822 | 0.640 | 0.754 | 0.973 | 0.853 | 0.731 |
| 14:30(8) | 0.642 | 0.750 | 0.546 | 0.705 | 0.610 | 0.640 | 0.928 | 0.856 | 0.921 | 0.678 |
| 14:45(9) | 0.679 | 0.544 | 0.538 | 0.606 | 0.886 | 0.843 | 0.579 | 0.734 | 0.573 | 0.891 |

Fig. 1: Decoding accuracy in the dynamic environment.

GSense [16] embeds symbols into gaps between customized packet preambles. B2W2 [17] mimics the DAFSK for communication from BLE to WiFi. C-Morse [18] constructs the radio energy patterns with Morse Coding. WiZig [6] improves the throughput by using multiple amplitudes. FreeBee [4] leverages the transmitting timing of beacon packets as the information carrier. DCTC [19] utilizes the transmitting timing of application packets to encode data. Another information carrier is the channel state. ZigFi [8] uses the impacts of ZigBee packets on Channel State Information (CSI) to convey data. Recently, researchers utilize physical layer information to achieve high-speed CTC. SymBee [9] creates distinguishing phase patterns on WiFi receiver with special ZigBee pay-load. WEBee [10] and BlueBee [11] emulate the signal of another technology in the payload. XBee [20] utilizes bit patterns to decode ZigBee packets at BLE. TiFi [21] utilizes backscattered harmonic to achieve CTC between WiFi and RFID. WIDE [12] emulates the phase shift of the receiver directly to achieve digital emulation. Meanwhile, CTC has many practical applications, such as channel coordination [22] and time synchronization [23].

Despite the tremendous advances, existing CTC approaches usually focus on enabling the communication between incompatible technologies. However, how to maintain the reliable performance in the intrinsically dynamic channels has not received enough attention. Existing methods only reactively bear the channel dynamics and heuristically sacrifice performance of throughput to lower the SER by retransmission [4], [10], increasing symbol window length [6], [7], controlling transmission power [8]. Different from existing methods, we directly include the channel state into our decoding model and exploit online learning to continuously adapt to the channel dynamics.

## III. MOTIVATION

In this section, we study the performance of existing CTC techniques in dynamic environments and further investigate the challenges that CTC encounters in dynamic environments.

Existing CTC methods usually use the threshold or the machine learning model as the decoding model to decode CTC symbols. Without losing generality, we investigate the performance of ZigFi [8] (a CTC from ZigBee to WiFi) as the example to show the impacts of dynamic environment on CTC. The basic idea of ZigFi is using the presence and absence of ZigBee packets to modulate the overlapped channel to transmit symbol 1 and 0. A WiFi receiver monitors the changes of Channel State Information (CSI) to detect the existence of ZigBee packet and then outputs the corresponding symbols after decoding. To detect the impacts of ZigBee transmissions on CSI, ZigFi defines two CSI features, variance and peak-to-peak (i.g. maximum - minimum), and decides whether there is ZigBee transmission by a binary SVM classifier.

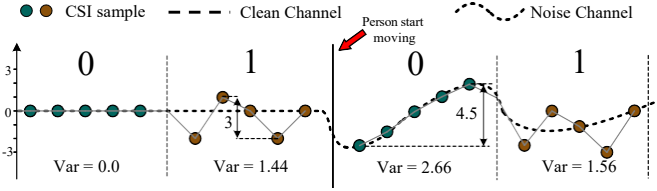

Fig. 2: The illustration of feature invalidation when the environment has low-frequency noise.

We conduct experiments in an office on our campus during working hours, to investigate the performance of ZigFi. We collect CSI sequences during the presences and absences of ZigBee packets and label them as symbol 1 and 0, respectively, as the labeled training data. The CSI sampling frequency is set to 2KHz and the collection of CSI sequence for each label lasts for 30 seconds. We continuously construct 10 training datasets at intervals of 15 minutes, and denote them as $D_0, D_1, ..., D_9$ in time order.

For each training dataset $D_i$, we train a decoding model from scratch only using $D_i$. Then we examine the performance of each decoding model. The test results are shown in Fig. 1. Each cell $A(i, j)$ denotes the decoding accuracy of the model trained by dataset $D_i$ and tested on dataset $D_j$. When we evaluate $A(i, i)$, we leave 20% of dataset $D_i$ for testing. From the results, we find that the decoding accuracy of all $A(i, i)$ is around 0.9, consistent with the reported results in ZigFi. However, the accuracy usually suffers from different degradations when testing the model on other slots ($i = j$), and the accuracy can decrease by more than 50% in one hour.

The results demonstrate that channel dynamics can cause significant performance degradation. The PHY information like CSI is much sensitive to channel dynamics and can be impacted by factors including multipath, scattering, fading, and power decay, besides the ZigBee transmissions. Hence, the original decoding model trained at a time will mismatch the current distribution of CSI features. The outdated model leads to performance degradation.

Besides, the inappropriate features can also result in performance degradation. Most of existing methods empirically predetermine a set of features, which may cause inaccuracy to the decoding model in dynamic channel environments. As shown in Fig. 2, the ZigBee transmissions fluctuate CSI in the clean channel. Hence, variance of the CSI, as a statistical feature, is used in ZigFi to distinguish the presence and absence of ZigBee transmissions. However, the channel dynamics can also impact the CSI. According to our measurements, we find that the human movements influence the multipath and introduce the low-frequency noise into CSI [24], as shown in Fig. 2. In the noise channel, both the variance and peak-to-peak (maximum - minimum) of symbol 0 are larger than those of symbol 1, confusing the decoding model and causing decoding errors.

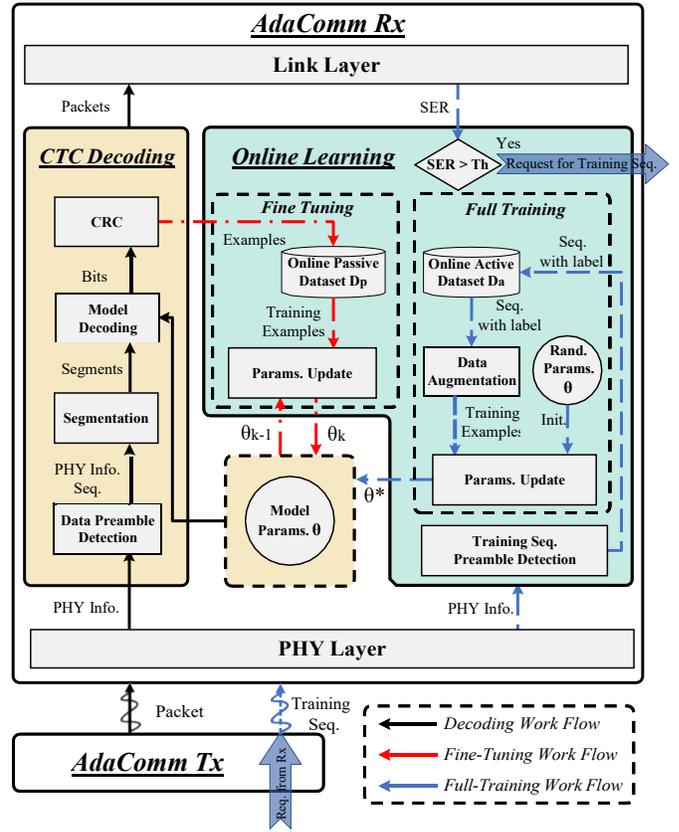

Fig. 3: The framework of AdaComm.

Actually, RSSI-based CTC such as WiZig also suffers from dynamic environments because the RSSI fluctuation will confuse the decoder that distinguishes symbols through the threshold of RSSI. Even though RSSI is less sensitive than CSI, it can still be influenced by both small-scale and large-scale channel dynamics. The small-scale dynamics are usually caused by multipath fading. The large-scale dynamics are usually caused by path loss via distance attenuation and shadowing effect by moving obstacles. When the wireless channel dynamics cause RSSI fluctuations, the number of decoding errors of RSSI-based CTC can significantly increase.

From the analysis and results, we can find that channel dynamics lead to not only the mismatch between features (e.g., CSI variation and RSSI amplitude) and the decoding models, but also invalidation of predetermined features. Existing feature-based methods cannot track and adapt to the real-time channel state with predetermined features and thus experience significant performance degradation.

## IV. DESIGN OF ADACOMM

AdaComm is a general platform for packet-level CTC. We propose an online learning mechanism to update the decoding model incrementally using online dataset, thus coping with the feature distortions. Meanwhile, we introduce a lightweight decoding model to automatically learn effective features from the raw data that are embedded with the information of current channel state.

### A. System Overview

Fig. 3 shows the architecture of AdaComm. AdaComm consists of two major parts: CTC decoding and online learning. The online learning can be further divided into two components: fine tuning and full training. In CTC decoding component, the data preamble detection module first detects the existence of modulated CTC symbols and passes the modulated channel state sequences to the segmentation module. The segments are then decoded by our decoding model (Section IV-B). The demodulated bits will be checked by be Cyclic Redundancy Check (CRC). The packets that pass CRC check will be delivered to CTC link layer.

Instead of training a feature-based model from an offline training dataset, we propose an online learning mechanism to update the model incrementally using the online dataset, thus helping the model acclimate to dynamic environments. *Fine-Tuning* (Section IV-C) and *Full-Training* (Section IV-D) are two updating modes to cope with the gradual and abrupt channel dynamics respectively. By leveraging the correctly decoded CTC packets that pass the CRC, AdaComm can continuously obtain the labeled CTC symbol data and learn the current channel state. Then AdaComm utilizes fine tuning mechanism to update the parameters of our decoding model to adjust to the channel. However, a channel state can change significantly and cause serious performance degradation. We propose the full training mechanism to recover from the failed model. In full training mode, the AdaComm receiver will actively send the request of the bursty training sequences to the AdaComm sender. Then the sender broadcasts the training sequences. Note that when channel state significantly changes, the data preamble detection method may also fail detecting CTC preambles due to the invalid decoding model. Hence, we propose a dedicated preamble for training sequences. To reduce the interrupt time of CTC, we propose a novel data augmentation method to obtain sufficient training data with a limited size of training sequences.

### B. Decoding Model

As discussed in motivation (Section III), the channel dynamics can cause the distortion of the features as well as the invalid features. The reason is that even though the raw data actually contain rich channel information, the extraction of manually defined features loses the information of channel state. Hence, a natural requirement comes to our mind: Can we extract effective features for decoding model automatically? A popular technique, neural network, is promising to meet the requirement because neural networks have the ability of automatically learning the features from raw data without any prior domain knowledge. Therefore, we explore the neural networks to automatically learn the features that embed the channel dynamic information.

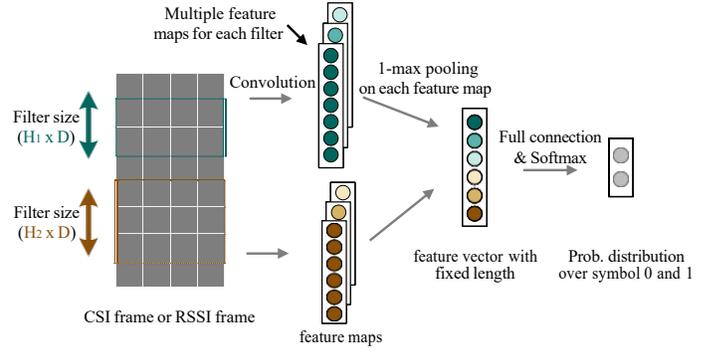

Fig. 4: Structure of our decoding model.

However, it is non-trivial to directly apply neural network techniques on CTC. Existing packet-level CTC techniques usually impact the transmissions from other technologies to convey data. For example, in ZigFi, the ZigBee CTC transmitter impacts the CSI on the WiFi receiver. But the original transmissions are not strictly periodic due to the use of CSMA/CD. Namely, the number of CSI samples is not constant in each CTC symbol window which is called a CSI frame. But traditional neural networks usually require fixed-dimension input and cannot directly process the CSI frame with a varying length. Simply interpolating or shrinking the CSI frames to a fixed size is infeasible because each CSI sample is independent, and directly interpolating samples may distort the channel information that we want to reserve.

To make full use of channel information embedded in the raw CSI frame, we propose a decoding model modified from Text-CNN [25]. Text-CNN is originally designed for short text classification in natural language processing and is able to deal with the varying-size inputs without interpolation. The architecture of our decoding model is shown in Fig. 4. Let $x_{i,j} \in \mathsf{R}^D$ denotes $j$-th sample in $i$-th CSI frame ($x_i \in \mathsf{R}^{N \times D}$), where $N$ is the frame length and $D$ is the number of overlapping subcarriers between WiFi channel and ZigBee channel (When this model is used to process RSSI, $D$ is set to 1). We apply convolution filters $w \in \mathsf{R}^{H \times D}$ on samples with a sliding window of $H$ to obtain new features. We set the step size of the sliding window on CSI frame as 1 and then each filter $w$ will produce a new feature vector with length $N - H + 1$. Then, 1-max pooling is applied over each feature vector to capture the most important feature. This pooling scheme naturally deals with variable CSI frame length, because it converts variable feature length ($N\_H +1$) to fixed length 1. Finally, this fixed length layer will pass through full connection and softmax layers to classify symbols.

### C. Fine Tuning

Existing CTC methods reactively adjust behaviors only when PER significantly increases. They fail to adjust in time to obtain the optimal CTC performance in dynamic channel environments. Hence, to tackle the gradual channel dynamics, AdaComm integrates the fine tuning mode that reuses the correctly decoded data to update the decoding model.

*1) Passive Data Collection:* Suppose we collect $n$ consecutive frames $x_0, x_1, ..., x_{n-1}$. The CTC data decoding model classifies the frames as $n$ symbols $b_0, b_1, ..., b_{n-1}$. If these $n$ symbols pass CRC, we obtain the new labeled data and append them to the passive dataset $D_p$, i.e., $D_p = D_p \cup \{(x_0, b_0), (x_1, b_1), ..., (x_{n-1}, b_{n-1})\}$. Note that, the whole process of collecting $D_p$ is non-intrusive. No extra cost of data collection is introduced. We can continuously accumulate the labeled data as long as there are new frames decoded correctly.

*2) Parameters Update:* After obtaining the labeled data, we can update the model parameters. Generally speaking, the training processes of the decoding model can be converted into solving the following optimization problem.

$$\min_{\theta \in \mathbb{R}^n} \mathbb{E}_{(x,y) \sim D}[J_\theta(x, y)] \quad (1)$$

where $x$ is CSI frame and $y$ is its label (e.g. symbol 0 or 1). $J$ is a loss function. There are many optimization methods that could be expressed as rules for updating its parameters $\theta$. For example, the iteration of Stochastic Gradient Descent (SGD) [26], a simple yet practical optimization method, can be described as

$$\theta_k = \theta_{k-1} - a \mathbb{E}[\nabla_{\theta_{k-1}} J(x, y)], \quad (x, y) \sim D_k, \quad (2)$$

where $\theta_k$ denotes the parameters of the decoding model in the $k$-th iteration. $a$ is the step size of updating. $\nabla_{\theta_k} J$ denotes the gradient with respect to $\theta_k$, and $D_k$ means $k$-th training batch. The training batch is sampled from the passive dataset $D_p$.

### D. Full Training

When wireless channel changes abruptly, there will be significant changes in underlying CSI features, which mismatch with features that decoding model has already learned. Fine tuning mode is disabled because the decoding model has already failed and no more CTC frames decoded correctly is available for passive dataset $D_p$. To cope with abrupt channel dynamics, we propose full training to quickly recover from the model failure.

In the full training mode of AdaComm, the receiver can detect the abrupt channel dynamic by monitoring the PER. If PER suddenly increases, the receiver will decide there is an abrupt change and initiate the full training mode. The receiver first sends the request of special training sequences to the sender. Once receiving the request, the sender will broadcast the training sequence. After receiving the training sequence, the receiver will train the decoding model from scratch to adapt the new channel environment. Since the original decoding model is out of date, the data preamble component may fail. Hence, we design a special training sequence with barker code based preamble to enable the detection of training sequence in a dynamic environment. We also propose a novel

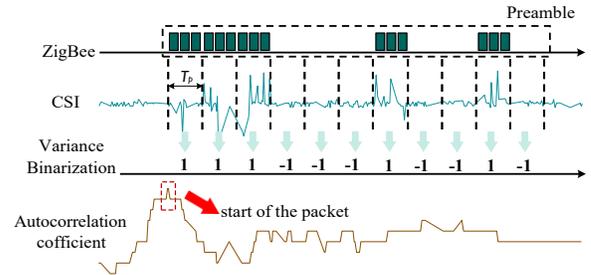

Fig. 5: Detection of the barker code based preamble.

data augmentation method to cut the cost of data collection, shortening the interruption time.

*1) Active Data Collection:* The training data for full training is collected from our training sequences. We construct the training sequence with a predefined pattern to help the receiver extract labeled data. Besides the preamble introduced in the next subsection, the training sequence consists of continuous symbol 1s followed by continuous symbol 0s. The periods for continuous symbol 1 and 0 are both $T_g$. Then we use the symbol window length $T_s$ to segment the sequence to obtain symbol frames $\{x_j\}$. The label $y_j$ of each frame can be inferred from the fixed structure of the training sequence. Then we have the online active dataset $D_a = \{(x_j, y_j)\}$.

*2) Training Sequence Preamble Detection:* Inaccurate decoding model may lead to inaccurate synchronization which will cause offsets when segmenting the labeled symbol frames, thus introducing extra noise to training dataset $D_a$. Hence, we need a more distinctive preamble to help the receiver reliably detect and locate the training sequence.

AdaComm leverages the property of barker code to design the preamble of training sequence for detection and synchronization. Barker code is the subset of PN sequences, and its key property is the ideal autocorrelation. Consider a barker code with length N, for all $1 \leq v < N$, the autocorrelation coefficient can be expressed as $c_v = \sum_{j=1}^{N-v} b_j b_{j+v}$, where $b_j$ is an individual barker code taking value between $+1$ and $-1$, for $1 \leq j \leq N$. The autocorrelation coefficient $c_v$ will reach its peak value when $v = 0$. In AdaComm, we adopt a 11-chip barker code, which is $\{+1, +1, +1, -1, -1, -1, +1, -1, -1, +1, -1\}$. We attach the preamble at both the head and the tail of the training sequence to improve the reliability.

On the receiver side, we continuously collect CSI samples and monitor the arrival of preambles. The detection process is shown in Fig. 5. We first convert the raw CSI into $+1$ or $-1$. We leverage the variance of CSI to decide the segment should be $+1$ or $-1$ according to a decision threshold, $Th$. If the variance of CSI in a window is larger than $Th$, the window is regarded as $+1$. Otherwise, the window is $-1$. We then calculate the autocorrelation coefficient $c_v$ using these 11 outputs. If the peak of $c_v$ exceeds a threshold, we assert there is an arrival of the preamble and the index of the corresponding CSI sample is the start of the training sequence. Even though we simply use the variance of CSI for binarization, our method is robust to detect the training preamble for two reasons. First, the symbol window length $T_p$ for preamble is twice as long as the data symbol $T_s$, providing much more redundancy. Second, the barker code with 11 chips has a fault-tolerant capability. The robustness of our training sequence preamble detection is evaluated in Section V-E.

*3) Data Augmentation:* It is known that more examples are fed to the model, more accurate the model tends to be. However, in CTC communication, the cost of actively collecting sufficient labeled data is unacceptable. First, the active collection impedes the normal CTC, leading to interruption in CTC service. Long-period collection can cause too long service interruption time. Second, the energy cost of ZigBee nodes during active data collection can be high because the radio has to keep active for a long period. To quickly obtain sufficient training data at low cost is crucial for the full training.

To tackle the dilemma, we devise a novel data augmentation method to generate emulated training data from the collected data. The basic principle of data augmentation is that semantic information of original data is still preserved after augmentation. So we can directly label the new generated data with the label of the original data.

We design our CSI data augmentation based on the observation that impacted CSI samples are randomly distributed. In fact, the impacts of ZigBee transmissions on CSI samples rely on the collisions between ZigBee packets and the WiFi preamble. Due to the asynchronous transmissions of ZigBee and WiFi packets, even though the impacts exist, the positions of impacted CSI samples are uncertain. Hence, using the collected CSI samples as the seeds for generation, random permutation can create much more labeled data while still maintaining channel information, including both the impacts of ZigBee transmissions and the channel dynamics.

Suppose the online dataset we actively collect is $D_a = (x_i, y_i)$, where $(x_i, y_i)$ is a pair of a CSI frame and the corresponding label $y_i$ (symbol 0 or symbol 1). For each pair $x_i, y_i$ in $D_a$, we perform random permutation on $x_i$ and get a new frame $x_i^j$. The only difference between $x_i$ and $x_i^j$ is the order of CSI samples, while the major semantic information of $x_i$ reserves. So the pair $(x_i^j, y_i)$ can be used as a new training example.

The full training module exploits the same optimization method as fine tuning but with the larger step size $\alpha$ and of course different training datasets. After random initialization of $\theta$, we feed new generated examples to the model batch by batch until the desired accuracy is achieved or timeout. To speed up training process, we improve generalization of model by discarding the generated examples that have already been fed to the model before.

### E. Generality of AdaComm

AdaComm is a general CTC framework that leverages online learning to adapt to the channel dynamics and obtains the reliable performance in dynamic environments. Although we use ZigFi, a CSI-based CTC, as the example to present our design details, we also apply AdaComm to RSSI-based CTC, WiZig [6] in our implementation. The main difference is using raw RSSI samples as input of decoding model. The CTC techniques with RSSI as information carrier leverage the threshold to decode symbols, which can be regarded as feature-based decoding model as well. They also have the problem of performance degradation due to small-scale fading and large-scale fading in dynamic wireless channel. Our evaluation shows that applying AdaComm to both CSI-based CTC and RSSI-based CTC indeed brings benefits to reliable performance.

## V. EVALUATION

In this section, we extensively evaluate AdaComm in various scenarios. We first introduce the experiment setting, and then present the performance and cost of AdaComm, compared with ZigFi (a CSI-based CTC from ZigBee to WiFi) and WiZig (a RSSI-based CTC from WiFi to ZigBee).

### A. Experiment Setup

We implement ZigFi, WiZig, CSI-based and RSSI-based AdaComm to study the performance improvement of AdaComm. For ZigFi and CSI-based AdaComm, two computers with Intel 5300 NICs act as WiFi sender and WiFi receiver to build the WiFi link. We use CSITool [27] to collect the CSI samples with an average sampling frequency of 2KHz. We implement the ZigBee sender on TelosB, a commercial ZigBee platform. The CTC symbol length $T_s$ is set to 4ms, the default setting in ZigFi. For WiZig and RSSI-based AdaComm, an USRP/N210 device acts as the CTC transmitter and a TelosB node acts as CTC receiver. The sampling frequency of RSSI is 1KHz. Unless otherwise specified, the communication channel of WiFi and ZigBee is set to channel 11 and channel 23 respectively to overlap in the frequency domain. To generate controllable channel dynamics, we also use another USRP/N210 to simulate Additive White Gaussian Noise (AWGN). Due to the limited space, in the following, we mainly present the comparison results of ZigFi and CSI-based AdaComm because the comparisons between WiZig and RSSI-based AdaComm show similar performance trends and conclusion.

### B. Performance Comparison under Different Settings

We first evaluate AdaComm under various settings, including the intensity of dynamics, transmission distance, and transmission power. We implement the default ZigFi that uses a pre-trained decoding model. We also implement an improved version of ZigFi with online learning mechanism. The performance metrics are throughput and SER.

*1) Impacts of Environment Dynamics:* We first evaluate AdaComm in different environments. We intentionally control the environment to construct the dynamics with different intensities. We have four scenarios: (1) relative static environment with no human and device movement; (2) a person keeps walking around; (3) the WiFi Tx keeps moving; (4) the WiFi

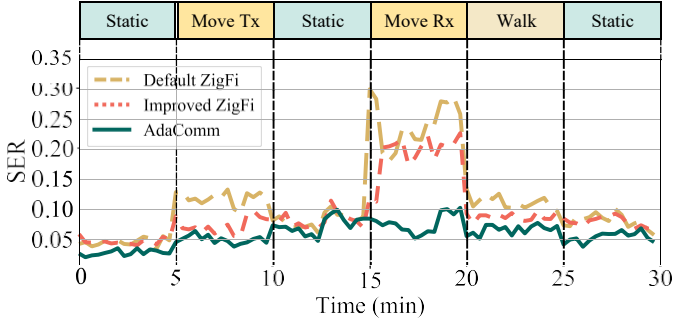

Fig. 6: SER in dynamic environments.

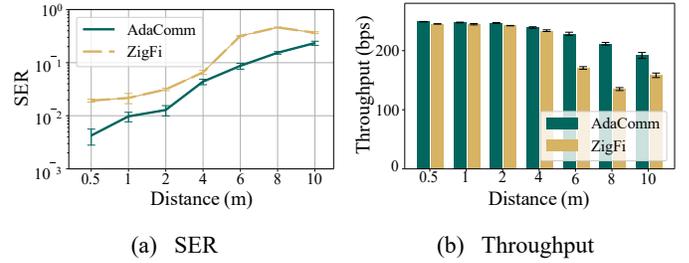

Fig. 7: SER and throughput vs. distance.

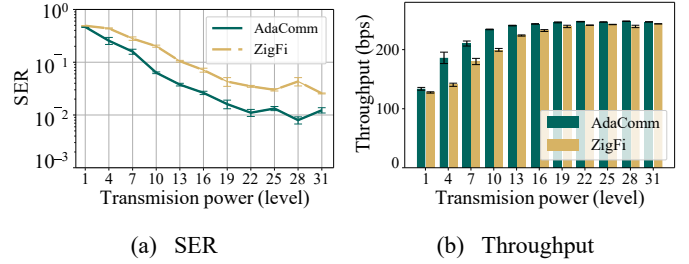

Fig. 8: SER and throughput vs. transmission power.

Rx keeps moving. The impacts on the decoding model of these four scenarios are expected to be more and more serious because the moving Rx directly impacts the received signals. We construct an environment varying sequence, as shown in the Fig. 6. Each scenario lasts for 5 minutes. Then we record the SER of AdaComm, default ZigFi, and improved ZigFi during the experiments.

Fig. 6 shows the SER of three methods during the varying environments. In the static environments, all three methods can achieve the SER lower than 0.1. However, when we move the WiFi Tx or ask a person to walk around, the SER of default ZigFi increases sharply, while AdaComm can quick adapt to this abrupt change with the help of online learning. Also benefiting from online learning, improved ZigFi achieves a relatively low SER. However, its SER increases from the 8 minutes because predefined features suffer from distortion, while AdaComm can keep the statable and low SER.

When we move the WiFi Rx, the channel experiences severe dynamics, the SER of default ZigFi sharply increases to more than 0.25. Even though we retrain the decoding model in improved ZigFi, the SER only keeps low for a very short time (less than 30s) and then increases to a very high level. This is because the fast changing environment can cause the failure of the retrained model. We can see that AdaComm maintains a quite low SER for all the time. The average SERs of AdaComm in four scenarios are 0.051, 0.066, 0.049, and 0.076 respectively. The SER variation is very limited. The results demonstrate that with online learning, AdaComm can deal with the sudden and gradual channel dynamics and achieve the reliable CTC with a quite low and stable SER.

*2) Impacts of Distance:* The distance between the ZigBee sender and the WiFi receiver affects the ZigBee's signal strength as well as the CSI variations at the receiver. We vary the distance from 0.5m to 10 meters to investigate the impacts of distance on the performance of AdaComm and ZigFi. The distance of the WiFi receiver and WiFi sender is fixed at 3m. The transmission power of ZigBee is set as level 16 (-6 dBm).

Fig. 7(a) and 7(b) present SER and throughput, respectively. As expected, both AdaComm and ZigFi have performance degradation with the increase of distance. When distance is 6m, the SER of ZigFi is 0.317, exceeding the required SER, 0.1. The SER of AdaComm is 0.086 which is 72.9% smaller than ZigFi. The corresponding throughput of AdaComm and ZigFi are 229 bps and 171 bps respectively. The reason that AdaComm outperforms ZigFi is that channel fading due to long distance can also cause the insatiability of features, which can be regarded as a kind of channel dynamics. The automatically learned features in AdaComm can distinguish symbols better than the predetermined features in ZigFi decoding model.

*3) Impacts of Transmission Power:* Besides distance, the transmission power of ZigBee also impacts the signal strength at the WiFi receiver. We vary the transmission power of ZigBee from power level 1 (-24dBm) to level 31 (0dBm), to study the performance of AdaComm and ZigFi. Both the distance between the ZigBee sender and the WiFi receiver and the distance between the WiFi sender and receiver are 3m.

Fig. 8 presents the results. We observe the similar results, compared with the results when varying distance. When increasing the transmission power, the received signal strength becomes stronger and the features become more distinctive. Hence, the SER decreases and the throughput increases. But the performance gap between AdaComm and ZigFi still exists due to the online learning based adaption in AdaComm.

### C. Communication Interrupt Time

In dynamic channel environments, CTC can encounter communication interrupt when the CTC cannot provide the communication service with the desired SER. To provide reliable service in practice, the communication interrupt time should be as short as possible. In our online updating framework, the total communication interrupt time $T_I$ consists three major parts: the data collection time $T_{collect}$, the training time $T_{train}$, and the model failure time $T_{failure}$. During model failure time, CTC may continue the CTC transmission but the achieved SER is higher than the required SER.

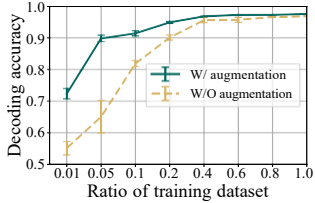 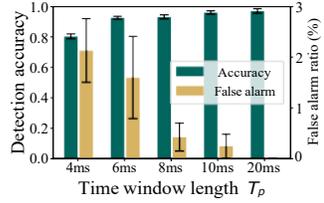 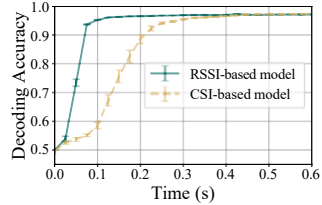 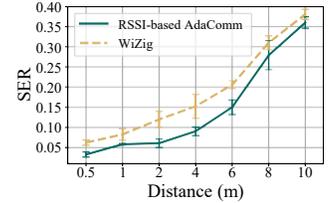

(a) Interrupt time   (b) Collection time   (c) Training time   (d) Fault time

Fig. 9: Communication interrupt time in different dynamic environments.

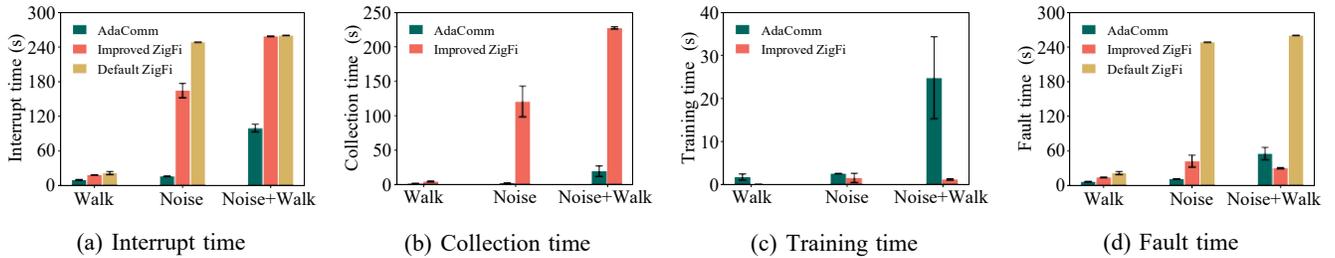

Fig. 10: Decoding accuracy of models trained on different ratios of dataset.

Fig. 11: Detection accuracy and false alarm ratio of preamble detection.

Fig. 12: Convergence time of the RSSI-based and the CSI-based decoding model.

Fig. 13: Performance improvement of RSSI-based AdaComm.

We control the channel environment to form three scenarios: (S1) a person keeps walking around the WiFi sender and receiver to generate the low-frequency noise; (S2) A USRP generates Gaussian noise with a high frequency; (S3) including the noise both in S1 and S2. The experiment in each scenario lasts for 5 minutes. We compare the interrupt time of default ZigFi, improved ZigFi, and AdaComm.

Fig. 9 presents the total communication interrupt time and the detailed compositions of the three CTC methods. We can find that $T_I$ of improved ZigFi and AdaComm is shorter than default ZigFi in the walking scenario and noise scenario. This is because the online updating used in improved ZigFi and AdaComm help to adjust the decoding model to overcome the channel dynamics. But when the environment is becoming complicated, simply updating the SVM parameters in improved ZigFi cannot handle and the performance degrades. Due to the retraining of models, improved ZigFi and AdaComm have additional data collection cost and training cost, as shown in Fig. 9(b) and Fig. 9(c). Due to the data augmentation, AdaComm can significantly reduce the data collection cost, as shown in Fig. 9(b). But the training time of AdaComm is longer than improved ZigFi because the Text-CNN has a more complex structure than SVM. Fig. 9(d) shows the model failure time. One interesting observation is that the model failure time of improved ZigFi is shorter than AdaComm in S3. This is because SVM cannot deal with the complicated channel dynamics and even retraining cannot provide satisfied SER. Hence, the most time is spent on data collection, as shown in Fig. 9(b).

### D. Cost Reduction of Data Augmentation

Next, we evaluate the contribution of data augmentation to reducing data collection cost. We continuously collect CSI samples for 60 seconds. The first 30s data form the complete training set and the last 30s data form the testing set. Then we vary the size of training set by controlling the ratio of used data in the complete training set. We vary the ratio from 1% to 100%. We use the same testing set to test the accuracy of our decoding model. The results are shown in Fig. 10. Our data augmentation can significantly reduce the size of needed data, to achieve the similar accuracy. Without data augmentation, achieving the accuracy of 0.9 needs more than 6s. With the help of data augmentation, AdaComm only needs 5% of the complete 30s training set which takes 1.5s to collect data.

### E. Accuracy of Preamble Detection

Barker code based preamble is used for training sequence detection and synchronization. The detection accuracy is highly related to the preamble symbol window length $T_p$. Hence, we vary $T_p$ and study the performance of our preamble detection. We measure two metrics: the accuracy of preamble detection and false alarm ratio. The results are shown in Fig. 11. As expected, when we increase $T_p$, the accuracy increases and false alarm ratio decreases because the binarization becomes more robust. When $T_p = 8ms$, the accuracy is 0.932 and false alarm ratio is just 0.42%. Hence, we set $T_p$ to $8ms$ in our current implementation.

### F. Convergence Time of Model Training

We investigate the convergence time of our model during full training. We run the training algorithm on CPU of a common WiFi PC receiver (Intel Core i5 @ 2.3GHz). Fig. 12 shows the accuracy of the RSSI-based and the CSI-based model changing over time. The accuracy of the CSI-based model reaches 0.9 within 0.2s and the training converges after 0.4s. The accuracy of the RSSI-based model reaches 0.93 within 0.075s and the training converges after 0.1s.

The fast convergence rate is because our decoding model adopts a lightweight neural network structure. Different from deep neural networks with millions of trainable parameters, our decoding model is quite lightweight. The number of parameters of decoding model is just 9794 (38.3KB) and 1346 (5.3KB) for CSI-based and RSSI-based AdaComm, respectively. The overhead is affordable for even low-power devices.

*G. Performance Improvement for RSSI-based CTC*

Due to large-scale fading, RSSI will also fluctuate when the distance between WiFi and ZigBee varies. Existing threshold-based decoding method is expected to have performance degradation when RSSI fluctuates. We vary the distance between the WiFi CTC sender and the ZigBee CTC receiver from 0.5m to 10 meters to investigate the impacts of distance on the performance of AdaComm and WiZig. The transmission power gain is fixed to 0 dB. Fig. 13 presents the results. As expected, when the distance increases, the SERs of both methods increase because of path loss. When distance is 2m, the SER of AdaComm is 0.061 which is 49.2% smaller than WiZig. Compared with the results of CSI-based AdaComm, we can find that even though AdaComm can improve the CTC performance for both CSI-based and RSSI-based CTC approaches, AdaComm has more significant performance improvement for CSI-based CTC. This is because CSI is more sensitive to the channel dynamics and the decoding features of existing predetermined decoding models are thus easily corrupted.

## VI. CONCLUSION

In this paper, we propose AdaComm, a general online learning CTC framework that maintains reliable performance in dynamic channel environments. AdaComm can automatically learn the effective features directly from the raw data that contains the information of current channel state. AdaComm utilizes two online learning modes, fine tuning and full training, to cope with the continuous and abrupt channel dynamics, respectively. In fine tuning, AdaComm keeps updating the receiver's decoding model by the correctly decoded CTC data. To quickly recover from the model failure caused by significant channel changes, we also propose the full training mode with data augmentation to obtain the new decoding model with only a limited overhead. We implement AdaComm and evaluate its performance in various environments. The experimental results show that AdaComm can reduce the SER by up to 72.9% and 49.2% for the CSI-based and the RSSI-based CTC respectively.

ACKNOWLEDGMENT

This work was supported by National Key R&D Program of China No.2017YFB1003000, National Natural Science Foundation of China No. 61772306, No.61672372 and No. 61672320.